# A Metric of Software Size as a Tool for IT Governance

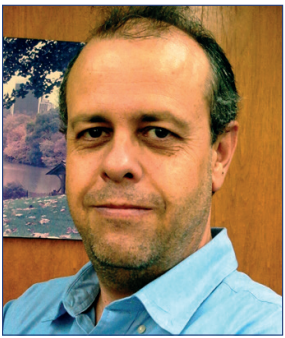

**Marcus Vinícius Borela de Castro**

is a public servant of the Federal Court of Accounts in Brazil, graduated in Computer Science from the Federal University of Viçosa, with a specialist degree in Information Technology Governance from the University of Brasilia and certified in metric function points (CFPS).

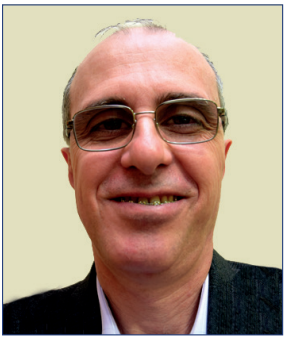

**Carlos Alberto Mamede Hernandes**

is a servant of the Federal Court of Accounts in Brazil, graduated in Data Processing from the University of Brasilia and with a Master's Degree in Knowledge Management and Information Technology at the Catholic University of Brasilia.Católica de Brasília.

## ABSTRACT

This paper[1] proposes a new metric for software functional size, which is derived from Function Point Analysis (FPA), but overcomes some of its known deficiencies. The statistical results show that the new metric, Functional Elements (EF), and its submetric, Functional Elements of Transaction (EFt), have higher correlation with the effort in software development than FPA in the context of the analyzed data. The paper illustrates the application of the new metric as a tool to improve IT governance specifically in assessment, monitoring, and giving directions to the software development area.

## 1. INTRODUCTION

### 1.1 RESEARCH SUBJECT

Organizations need to leverage their investments in technology to create new opportunities and produce change in their capabilities [RUBIN 1993, p. 473]. According to ITGI [2007, p. 7], information technology (IT) has become an integral part of business for many companies with key role in supporting and promoting their growth. In this context, IT governance fulfills an important role of directing and boosting IT in order to achieve its goals aligned with the company's strategy.

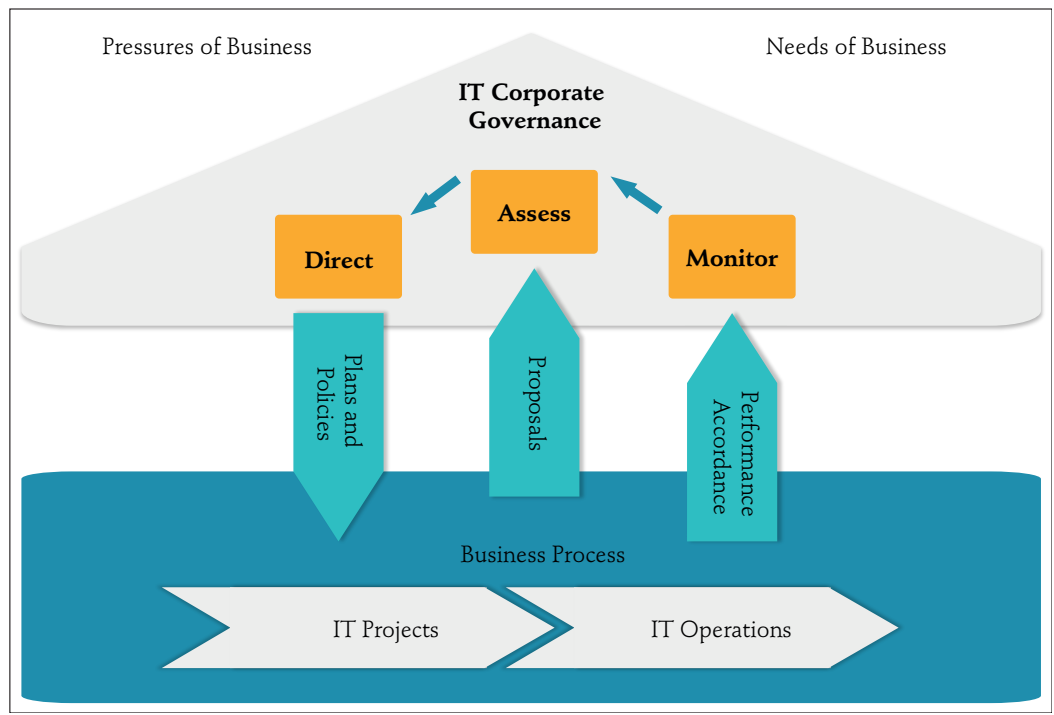

**Figure 1:**
Cycle Assess-Direct-Monitor of IT Governance

Source: ISO 38500 [3, p. 7]

In order for IT governance to foster the success of IT and of the organization, ISO 38500 [ISO, IEC, 2008, p. 7] proposes three main activities: to assess the current and future use of IT; to direct the preparation and implementation of plans and policies to ensure that IT achieves organizational goals; to monitor performance and compliance with those policies (Fig. 1).

A metric of software size can compose several indicators to help reveal the real situation of the systems development area for the senior management of an organization, directly or through IT governance structures (e.g., IT steering committee). Measures such as the production of software in a period (e.g., measure of software size per month) and the productivity of an area (e.g., measure of software size per effort) are examples of indicators that can support the three activities of governance proposed by ISO 38500.

For the formation of these indicators, one can use Function Point Analysis (FPA) to get function points (FP) as a metric of software size. Created by Albrecht [1979], FPA has become an international standard for measuring the functional size of a software with the ISO 20926 [ISO; IEC, 2009] designation. Its rules are maintained and enhanced by a nonprofit international group of users called International Function Point Users Group (IFPUG), responsible for publishing the Counting Practices Manual (CPM), now in version 4.3.1 [IFPUG, 2010].

Because it has a direct correlation with the effort expended in software development [ALBRECHT; GAFFNEY, 1983; KEMERER, 1987], FPA has been used as a tool for information technology management, not only in Brazil but worldwide. As identified in the *Quality Research in Brazilian Software Industry report*, 2009 [BRASIL, 2009, p. 93], FPA is the most widely used metric to evaluate the size of software among software companies in Brazil, used by 34.5% of the companies. According to a survey carried out by Bundschuh and Dekkers [2008, p. 393], 80% of projects registered on the International Software Benchmarking Standards Group (ISBSG), release 10, which applied metric used the FPA.

The FPA metric is considered a highly effective instrument to measure contracts [VAZQUEZ et al., 2011, p. 191]. However, it has the limitation of not treating non-functional requirements[2], such as quality criteria and response-time constraints. Brazilian federal government institutions also use FPA for procurement of development and maintenance of systems. In addition to the several decisions[3] by the Brazilian Federal Court of Accounts (TCU) that point out FPA as an example of metric to be used in contracts. the *Metrics Roadmap* of SISP [BRASIL, 2012], a federal manual for software procurement, recommends its application to federal agencies.

Despite the extensive use of the FPA metric, a large number of criticism about its validity and applicability, described in Section 2.2, put in doubt the correctness of its use in contracts and the reliability of its application as a tool for IT management and IT governance.

So the question arises for the research: is it possible to propose a metric for software development, with the acceptance and practicality of FPA, that is, based on its concepts already widely known, without some of the flaws identified in order to maximize its use as a tool for IT governance, focusing on systems development and maintenance?

## 1.2 RESEARCH RATIONALE

The rationale for the research can be analyzed based on the interest of several players involved in the context of software development and maintenance:

1. governance committees: metric can derive indicators that will enable greater IT governability;

2. IT manager[4]: metric can allow for better control towards achieving the goals set by upper administration;

3. suppliers from the private market and public agencies: metric can increase objectivity of the relationship, enabling contracts with a lower probability of causing problems, with payment per results and fair prices;

4. oversight bodies, such as TCU: the metric can support evaluation, on a more objective basis, of software development public contracts (e.g. evaluations on IT planning, contract planning and contract management);

5. research institutions: the study proposed can serve as the foundation for new studies, after all, the field of metrics does not have many good research papers. A survey by Jörgensen and Shepperd (2007, p. 36) shows that most of the research of software costs does not take into consideration articles that have already been published and criticizes the obsolescence of the data used.

## 1.3 CLASSIFICATION OF METHODOLOGY

This paper can be classified as a practical research, according to the Demo classification (*apud* ANDRA-DE, 2002, p.4) since it aims at solving problems related to actual application, as mentioned in section 1.2.

According to Andrade (2012, p. 5-6), with regard to the objectives, the article is exploratory because it proposes a new approach to metric. It is also descriptive because it presents concepts, such as software metric, and illustrates its application in IT governance. As for the approach to achieve the objectives, the paper can be classified as deductive (ANDRADE, 2002, p. 11) since it proposes a new metric based on theoretical concepts. As for the procedures adopted, the paper uses the statistical method (ANDRADE, 2002, p. 14) to build and evaluate the results.

## 1.4 SPECIFIC OBJECTIVES

The specific objectives of this paper are:

1. to present an overview of software metrics and FPA;

2. to present the criticisms to the FPA technique that motivated the proposal of a new metric;

3. to derive a new metric based on FPA;

4. to evaluate the new metric against FPA in the correlation with effort;

5. to illustrate the use of the proposed metric in IT governance in the context of systems development and maintenance.

Each of the specific objectives is dealt with in its own subsection under the topic Development below.

## 2. DEVELOPMENT

## 2.1 SOFTWARE METRICS

### 2.1.1 Conceptualization, categorization, and application

Bundschuh and Dekkers [2008, p. 180-181] describe various interpretations for metric, measure, and indicator found in the literature. Concerning this study, no distinction is made among these three terms. We used Fenton and Pfleeger's definition [1988, p. 5] for measure: a number or symbol that characterizes an attribute of a real world entity, object or event, from formally defined rules[5].

| | Criterion | Category | Source |
|---|---|---|---|
| Table I: Examples Of Categories Of Software Metrics | Entity | Process<br>Product<br>Resource | [13, p. 74] |
| | Number of attributes involved | Direct<br>Indirect | [13, p. 39] |
| | Target of differentiation | Size<br>Quality | [15, p. 32] |

According to Fenton and Pfleeger [1998, p. 74], software metrics can be applied to three types of entities: processes, products, and resources. The authors also differentiate direct metrics, when only one attribute of an entity is used, from indirect metrics, the other way around [FENTON; PFLEEGER, 1998, p. 39]. Indirect metrics are derived by rules based on other metrics. The speed of delivery of a team (entity type: resource) is an example of indirect metric because it is calculated from the ratio of two measures: size of developed software (product) development and elapsed time (process). The elapsed time is an example of direct metric. Moser [1996, p. 32] differentiates size metrics from quality metrics: size metrics distinguish between the smallest and the largest whereas quality metrics distinguish between good and bad. Table I consolidates the mentioned categories of software metrics.

Moser [1996, p.31] notes that, given the relationship between a product and the process that produced it, a product measure can be assigned to a process, and vice versa. For example, the percentage of effort in testing, which is a development process attribute, can be associated with the generated product as an indicator of its quality. Additionally, the number of errors in production in the first three months, a system attribute (product), can be associated to the development process as an indicative of its quality.

Fenton and Pfleeger [1998, p. 12] set three goals for software metric: to understand, to control, and to improve the targeted entity. They call our attention to the fact that the definition of the metrics to be used depends on the maturity level of the process being measured: the more mature, more visible, and therefore more measurable [FENTON; PFLEEGER, 1998, p. 83]. Chikofsky and Rubin [1999, p. 76] highlight that an initial measurement program for a development and maintenance area should cover five key dimensions that address core attributes for planning, controlling, and improvement of products and processes: size, effort, time, quality, and rework. The authors remind us that what matters are not the metric itself, but the decisions that will be taken from them, refuting the possibility of measuring without foreseeing the goal [CHIKOFSKY; RUBIN, 1999, p. 75].

According to Beyers [2002, p. 337], the use of metric in estimates (e.g., size, time, cost, effort, quality, and allocation of people) can help in decision making related to software development and to software projects planning.

### 2.1.2 FPA overview

According to the categorization in previous section, FPA is an indirect measure of product size. It measures the functional size of an application (system) as a gauge of the functionality requested and delivered to the user of the software. This is a metric understood by users, regardless of the technology used[6].

It is worth mentioning that, in addition to FPA, there are four other functional metrics considered ISO standard of functional metric since they follow the rules defined in the six norms of the series ISO 14143 (ISO; IEC, 2002a, 2003, 2004, 2006, 2007, 2011a): MKII FPA (ISO; IEC, 2002b), COSMIC-FFP (ISO; IEC, 2011b), FiSMA (ISO; IEC, 2010) e NESMA (ISO; IEC, 2005). According to Gencel and Demirors (2008, p.4), ISO standard functional metrics estimate software size based on the function delivered to users, with a difference in counted objects and in the way they are counted[7].

Functionalities can be of two types: transactions, which implement data exchanges with users and other systems, and data files, which indicate the structure of stored data. There are three types of transactions: external inquiry (EQ), external outputs (EO), and external inputs (EI), as the primary intent of the transaction is, respectively, a simple query, a more elaborate query (e.g., with calculated totals) or data update. There are two types of logical data files: internal logical files (ILF) and external interface files (EIF), as their data are, respectively, updated or just referenced (accessed) in the context of the application.

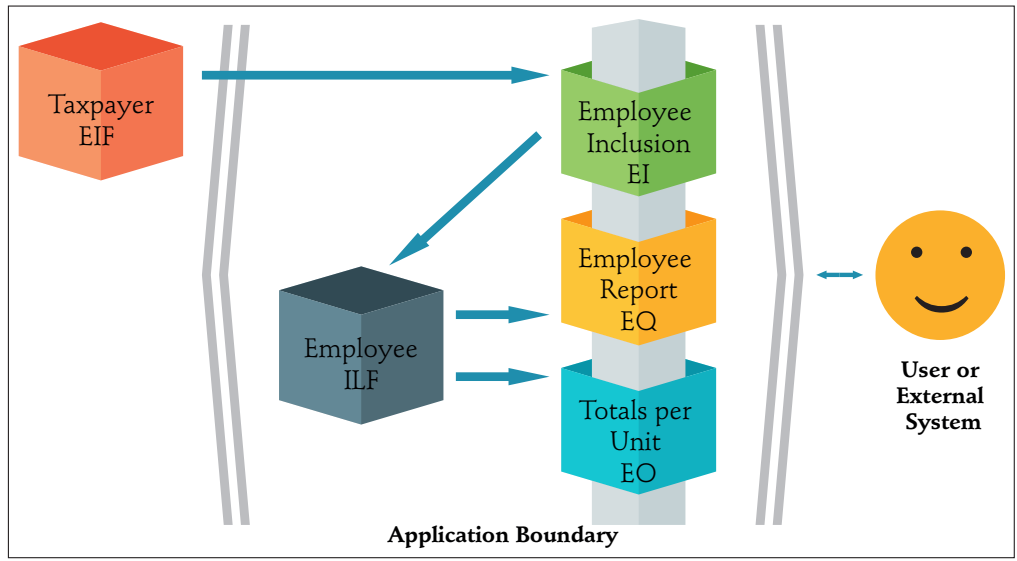

**Figure 2:**
Visualization of the five types of functions in FPA

Fig. 2 illustrates graphically these function types. To facilitate understanding, we can consider an example of EI as an employee inclusion form which includes information in the employees data file (ILF) and validates the tax code informed by the user accessing the external file taxpayers (EIF), external to the application, which contain Federal Revenue CPF data. Also in the application we could have, hypothetically, an employee report, a simple query containing the names of the employees of a given organizational unit (EQ) and a more complex report with the number of employees per unit (EO).

In the FPA calculating rule, each function is evaluated for its complexity and takes one of three classifications: low, medium or high complexity. Each level of complexity is associated with a size in function points.

Table II illustrates the derivation rule for external inquiries, according to the number of files accessed (File Type Referenced - FTR) and the number of fields that cross the boundary of the application (Data Element Type - DET).

As for EQ, each type of functionality (EO, EI, ILF, and EIF) has its specific rules for derivation of complexity and size, similar to Table II. Table III summarizes the categories of attributes used for calculating function points according to each type of functionality.

The software size is the sum of the sizes of its functionalities. This paper is not an in-depth presentation of concepts associated with FPA. Details can be obtained in the Counting Practices Manual, version 4.3.1 [IFPUG, 2010].

**Table II:**
Derivation Rule For Complexity And Size In Function Points Of An External Inquiry (Eq)

| FTR (file) \ DET (field) | 1 a 5 | 6 a 19 | 20 or more |
|---|---|---|---|
| 1 | low (3) | low (3) | medium (4) |
| 2 a 3 | low (3) | medium (4) | high (6) |
| 4 or more | medium (4) | high (6) | high (6) |

**Table III:**
Categories Of Functional Attributes For Each Type Of Functionality

| Function | Functional Attributes |
|---|---|
| Transactions: EQ, EO, EI | referenced files (FTR) and fields (DET) |
| Logical files: ILF, EIF | logical registers (Record Element Type - RET) and fields (DET)Campos (ou TD – tipos de dados) |

## 2.2 CRITICISMS TO THE FPA TECHNIQUE THAT MOTIVATED THE PROPOSAL OF A NEW METRIC

Despite the extensive use of the metric FPA, mentioned in Section I, there are a lot of criticism about its validity and applicability that call into question the correctness of its use in contracts and the reliability of its application as a tool for IT management and governance (ABRAN; ROBILLARD, 1994; FENTON; PFLEEGER, 1998; KITCHENHAM, 1997; KITCHENHAM; KÄNSÄLÄ, 1993; KITCHENHAM *et al.*, 1995; KRALJ *et al.*, 2005; PFLEEGER *et al.*, 1997; TURETKEN *et al.*, 2008; XIA *et al.*, 2009).

Several metrics have been proposed taking FPA as a basis for their derivation, either to adapt it to particular models, or to improve it, fixing some known bugs. To illustrate, there is Antoniol *et al.* [2003] work proposing a metric for object-oriented model and Kralj *et al.* [2005] work proposing a change in FPA to measure more accurately high complexity functions (item 4 below).

The objective of the metric proposed in this paper is not to solve all faults of FPA, but to help to reduce the following problems related to its definition:

1. low representation: the metric restricts the size of a function to only three possible values, according to its complexity (low, medium, or high). But there is no limit on the number of possible combinations of functional elements considered in calculating the complexity of a function in FPA;

2. functions with different functional complexities have the same size: as a consequence of the low representation. Pfleeger *et al.* [1997, p. 36] say that if H is a measure of size, and if A is greater than B, then $H_A$ should be greater than HB. Xia *et al.* [2009, p. 3] show examples of functions with different complexities that were improperly assigned the same value in function points because they fall into the same complexity classification;

3. abrupt transition between functional element ranges: Xia *et al.* [2009, p. 4] introduced this problem. They present two logical files, B and C, with apparent similar complexities, differing only in the number of fields: B has 19 fields and C has 20 fields. The two files are classified as low (7 fp, function points) and medium complexity (10 fp), respectively. The difference lies in the transition of the two ranges in the complexity derivation table: up to 19 fields, it is considered low complexity; from 20 fields, it is considered medium complexity. The addition of only one field leading to an increase in 3 pf is inconsistent, since varying from 1 to 19 fields does not involve any change in the function point size. A similar result occurs in other transitions of ranges;

4. limited sizing of high complexity functions: FPA sets an upper limit for the size of a function according to its type. Kralj *et al.* [2005, p. 83] describe the situation of functions that are improperly classified as being of high complexity. They call attention to the need to have higher numbers for greater complexities and propose a change in the calculation of FPA as a solution[8];

5. operation on ordinal scale: as previously seen, FPA involves classifying the complexity of functions in low, medium or high complexity, as a ordinal scale. These labels in the calculated process are substituted by numbers. An internal logical file, for example, receives 7, 10 or 15 function points, as its complexity is low, medium or high, respectively. Kitchenham [1997, p. 29] criticizes the inadequacy of adding up values of ordinal scale in FPA. He argues that it makes no sense to add the labels *low complexity* and *high complexity*, even if using labels 7 and 15 respectively as synonyms;

6. inability to measure changes in parts of the function: this characteristic, for example, does not allow to measure function points of part of a functionality that needs to be changed in one maintenance operation. Thus, a function addressed in several iterations in an agile method or other iterative process is always measured with full size, even if the change is considered small in each of them.

Given the deficiencies reported, the correlation between the size in function points of software and the effort required for the development tends not to be appropriate, since FPA has these deficiencies in the representation of the real functional size of software. If there are inaccuracies in the measuring of the size of what must be done, it is impossible to expect a proper definition of the effort and therefore accuracy in defining the cost of development and maintenance. The mentioned problems motivated the development of this work, in order to propose a quantitative metric, with infinite values, called Functional Elements (EF).

## 2.3 DERIVATION PROCESS OF THE NEW METRIC

The proposed metric, Functional Elements, adopts the same concepts of FPA but changes the mechanism to derive the size of function[9].

The reasoning process for deriving the new metric, as described in the following sections, implements linear regression similar to that seen in Graph 1. The objective is to derive a formula for calculating the number of EF for each type of function (Table VII in Section 2.3.4) from the number of functional attributes[10] considered in the derivation of its complexity, as indicated in Table II in Section 2.1.2.

The marked points in Graph 1 indicate the size in fp (Z axis) of an external inquiry derived from the number of files (X axis) and the number of fields (Y axis), which are the attributes used in the derivation of its complexity (see Table II in Section 2.1.2). The grid is the result of a linear regression of these points, and represents the value of the new metric.

### 2.3.1 Step 1 - definition of the constants

If the values associated with the two categories of functional attributes are zero, the EF metric assumes the value of a constant. Attributes can be assigned value zero, for example, in the case of maintenance limited to the algorithm of a function not involving changes in the number of fields and files involved. In the context of the new metric, the dimension of operation to exclude a functionality takes on the value of the constant, since there are no attributes specifically impacted by this operation.

The values assigned to these constants come from the NESMA, ISO standard of functional measurement, functional metric for the cases with zero-value attributes, as documented in *Function Point Analysis For Software Enhancement* (NESMA, 2009). FPA itself (IFPUG, 2010, v. 4, p. 94) indicates NESMA as an alternative for maintenance measurements due to its capacity to deal with the 6th criticism of section 2.2. NESMA scales maintenance by multiplication of the original size of the function by an impact factor of the alteration. The impact factor derives from the proportion between the volume of attributes (e.g. fields) included, altered or excluded and their original volume in the function. The adjustment factor takes on values that are multiples of 25%, up to the limit of 150%.

For each type of functionality, the proposed metric uses the smallest possible value by applying NESMA, that is, 25% of the number of fp of a low complexity function of each type: EIF - 1.25 (25% of 5); ILF - 1.75 (25% of 7); EQ - 0.75 (25% of 3); EI - 0.75 (25% of 3), and EO - 1 (25% of 4).

### 2.3.2 Step 2 - treatment of ranges with unlimited number of elements

In FPA, each type of function has its own table to derive the complexity of a function, in a similar way to Table II in Section II-A-2, which presents the values of the ranges of functional attributes for the derivation of the complexity of external inquiries. The third and last range of values of each functional element of the derivation tables of all types of functions is unlimited, as we see *20 or more* TD in the first cell of the fourth column of the same table, and *4 or more* ALR in the last cell of the first column.

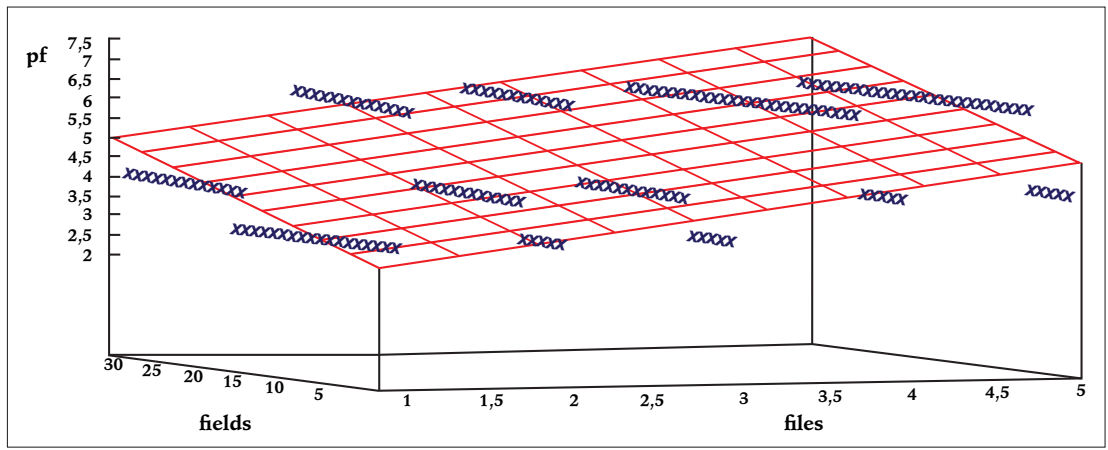

**Graph 1:** Derivation of number of fp of an external inquiry from the attributes used in the calculation

In order to create a finite set of data for regression, a superior limit was set for these ranges with a number of elements equivalent to that of the greatest precedent range[11]. In the case of ranges for external inquiries, the number of fields was limited to 33, a result of defining 14 as the number of elements of the third range (*20 to 33*), which is the same size of the largest range (*6 to 19 – 14 elements*). The number of referenced files was limited to 5, using the same reasoning. The limitation of the ranges is a mathematical artifice to prevent imposing an upper limit for the new metric (4th criticism in Section 2.2).

### 2.3.3 Step 3 - generation of points for regression

The objective of this phase was to generate, for each type of function, a set of data records with three values: the values of the functional attributes and the derived fp, already decreased from the value of the constant in step 1. Table IV illustrates some points generated for the external inquiry.

An application developed in MS Access generated a dataset with all possible points for the five types of functions, based on the tables of complexity with bounded ranges developed in the previous step. Table V shows all considered combinations of ranges for EQ.

### 2.3.4 Step 4 - linear regression

The several points obtained in the previous step were imported into Excel 2007 for linear regression between the size of FP and the functional attributes, using the Ordinary Least Squares Method (OLS) held constant with value zero, since these constants were already defined in step 1 and decreased from the expected value in step 3.

The statistical results of the regression are shown in Table VI for each type of function.

Table VII shows the derived formula for each type of function with coefficient values rounded to two decimal place values. Each formula calculates the number of functional elements, which is the proposed metric, based on the functional attributes impacting the calculation and the constants indicated in step 1. The acronym EFt and EFd represent the functional elements associated with transactions (EQ, EI, and EO) and data (ILF and EIF), respectively.

The functional elements metric, EF, is the sum of the functional elements transaction, EFT, with the functional elements of data, EFd, as explained in the formulas of Table VII. So the proposed metric is: EF = EFt + EFd.

The EFt submetric does not count logical files (ILF and EIF) in separate as in the EFd submetric, but only as they are referenced in the context of transactions. Files are also not counted in other ISO standard metrics of functional size [BUNDSCHUH; DEKKERS, 2008, p.

**Table IV:** Partial Extract Of The Dataset For External Inquiry

| FTR | DET | PF (decreased of constant of step 1) |
|---|---|---|
| 1 | 1 | 2,25 |
| 1 | 2 | 2,25 (…) |
| 1 | 33 | 3,25 |
| 2 | 1 | 2,25 (…) |

**Table V:** Combinations Of Ranges For Calculating Fp Of Eq

| Function type | Initial FTR | Final FTR | Initial DET | Final DET | Original FP | PF decreased of constant |
|---|---|---|---|---|---|---|
| EQ | 1 | 1 | 1 | 5 | 3 | 2,25 |
| EQ | 1 | 1 | 6 | 19 | 3 | 2,25 |
| EQ | 1 | 1 | 20 | 33 | 4 | 3,25 |
| EQ | 2 | 3 | 1 | 5 | 3 | 2,25 |
| EQ | 2 | 3 | 6 | 19 | 4 | 3,25 |
| EQ | 2 | 3 | 20 | 33 | 6 | 5,25 |
| EQ | 4 | 5 | 1 | 5 | 4 | 3,25 |
| EQ | 4 | 5 | 6 | 19 | 6 | 5,25 |
| EQ | 4 | 5 | 20 | 33 | 6 | 5,25 |

388]: MKII FPA [ISO; IEC, 2002b] and COSMIC-FFP [ISO; IEC, 2011b].

When evaluating the metric, in the next section, two of them were tested, EF and EFt, counting and not counting the logic files, and the results show that EFt has a better correlation with effort[12]. Although it was not assessed, submetric EFd has its worth because it reflects the structural complexity of the data of an application.

## 2.4 EVALUATION OF THE NEW METRIC

The new EF metric and its submetric EFt were evaluated for their correlation with effort in comparison to the FPA metric. The goal was not to evaluate the quality of these correlations, but to compare their ability to explain the effort[13].

We obtained a spreadsheet from a federal government agency with records of Service Orders (OS) contracted with private companies for coding and testing activities[14]. An OS contained one or more requests for maintenance or development of functions of one system, such as: create a report, change a transaction. The spreadsheet showed for each OS the real allocated effort and, for each request, the size of the function handled. The only fictitious data were the system IDs, functionality IDs and OS IDs, as they were not relevant to the scope of this paper. The spreadsheet showed the time spent in hours and the number of people allocated for each OS. The OS effort, in man-hours, was derived from the product of time by team size. Table VIII presents the structure of the received data.

**Table VI:** Statistical Regression - Comparing Results Per Types Of Functions

|  | ILF | EIF | EO | EI | EQ |
|---|---|---|---|---|---|
| $R^2$ | 0,96363 | 0,96261 | 0,95171 | 0,95664 | 0,96849 |
| Records | 729 | 729 | 198 | 130 | 165 |
| Coefficient *p-value* (FTR or RET) | 3,00E-212 | 1,17E-211 | 7,65E-57 | 1,70E-43 | 4,30E-60 |
| Coefficient *p-value* (DET) | 2,28E-231 | 2,71E-225 | 1,44E-59 | 2,76E-39 | 2,95E-45 |

**Table VII:** Calculation Formulas Of Functional Elements By Type Of Function

| Function type | Formula |
|---|---|
| ILF | $EFd = 1.75 + 0.96 * RET + 0.12 * DET$ |
| EIF | $EFd = 1.25 + 0.65 * RET + 0.08 * DET$ |
| EO | $EFt = 1.00 + 0.81 * FTR + 0.13 * DET$ |
| EI | $EFt = 0.75 + 0.91 * FTR + 0.13 * DET$ |
| EQ | $EFt = 0.75 + 0.76 * FTR + 0.10 * DET$ |

**Table VIII:** Structure Of The Received Data To Evaluate The Metric

| Abbreviation | Description | Domain |
|---|---|---|
| OS | Identification Number of a service order | up to 10 numbers |
| Function | Identification Number of a function | up to 10 numbers |
| Type | Type (categorization) of a functionality according to FPA | EQ, EI, EO, ILF or EIF |
| Operation | Operation performed, which may be inclusion (I) of a new feature or change (A) of a function (maintenance) | I or A |
| Final FTR RET | Value at the conclusion of the request implementation: if the function is a transaction, indicates the number of referenced logical files (FTR); if it is a logical file, indicates the number of logical records (RET) | up to 3 numbers |
| Operation FTR RET | Number of FTR or RET that were included, changed or deleted in the scope of a maintenance of a functionality (only in change operation) | up to 3 numbers |
| Original FTR RET | Number of FTR or RET originally found in the functionality (only in change operation) | up to 3 numbers |
| Final DET | Number of DET at the conclusion of the request implementation | up to 3 numbers |
| Operation DET | Number of DET included, changed or deleted in the scope of a functionality maintenance (only in change operation) | up to 3 numbers |
| Original TD | Number of DET originally found in a functionality (only in change operation) | up to 3 numbers |
| FP | Number of function points of the functionality at the conclusion of the request | up to 2 numbers |
| %Impact | Percentage of the original function impacted by the maintenance, as measured by NESMA [27] | 25, 50, 75, 100, 125, 150 |
| PM | Number of maintenance points of the functionality handled, as measured by NESMA [27] | up to 4 numbers |
| System | Identification of a system | one char |
| Hours | Hours dedicated by the team to implement the OS | up to 5 numbers |
| Team | Number of team members responsible for the implementation of the OS | up to 2 numbers |

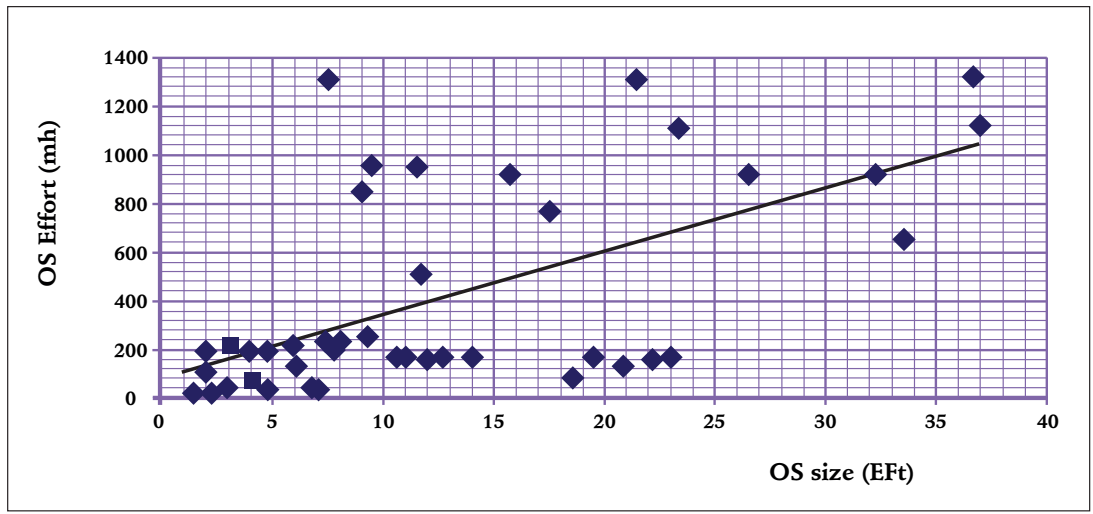

**Graph 2:**
Dispersion of points (OS) of H system: effort (man-hour) x size (Functional Element of Transaction)

Data from 183 Service Orders were obtained. However, 12 were discarded for having dubious information, for example, undefined values for function type, number of fields, and operation type. The remaining 171 service orders were related to 14 systems and involved 505 requests that dealt with 358 different functions. To achieve higher quality in the correlation with effort, we decided to consider only the four systems[15] associated with at least fifteen OS, namely, systems H, B, C, and D. Table IX indicates the number of OS and requests for each system selected.

The data were imported into Excel 2007 to perform the linear regression[16] using the ordinary least squares method after calculating the size in EF and EFt metrics for each request in an MS-Access application developed by the authors: between the effort and the size, calculated in the FP, EF and EFt metrics. The linear regression was carried out considering the constant with value zero, since there is no effort if there is no size[17]. The operation was done through a system because the variability of the factors that have an influence on effort are reduced within a single system[18]. Graph 2 illustrates the dispersion of points (OS) on the correlation between size and effort in EFt (man-hour) and the line derived by linear regression in the context of system H.

The coefficient of determination $R^2$ was used to represent the degree of correlation between effort and size calculated for each of the evaluated metrics. According to Sartoris [2008, p. 244], $R^2$ indicates, in a linear regression, the percentage of the variation of a dependent variable Y that is explained by the variation of a second independent variable X. Table IX shows the results of the linear regressions performed.

From the results presented on Table IX, comparing the correlation of the metrics with effort, we observed that:

1. correlations of the new metrics (EF, EFt) were considered significant at a confidence level of 95% for all systems (*p-value* less than 0.05[19]). However, the correlation of FPA was not significant for system B (*p-value* 0.088 > 0.05);

2. correlations of the new metrics were higher in both systems with the highest number of OS (H and B). A better result in larger samples is an advantage, because the larger the sample size, the greater the reliability of the results, since the *p-value* has reached the lowest values for these systems;

3. although no metric got a high coefficient of determination ($R^2 > 0.8$), the new metrics achieved medium correlation ($0.8 > R^2 > 0.5$) in the four systems evaluated, whereas FPA obtained weak correlation ($0.2 > R2$) in system B, considering the confidence level of 91.2% in this correlation (*p-value* 0.088);

4) correlations of the new metrics were superior[20] in three out of the four systems (H, B, and D), that is, in 75% of the systems.

**Table IX:** Results Of Linear Regressions - Effort Versus Metrics Of Size

| System | | H | B | C | D |
|---|---|---|---|---|---|
| Quantity of OS | | 45 | 25 | 21 | 15 |
| Quantity of Requests | | 245 | 44 | 60 | 20 |
| FP | $R^2$ | 59,3% | 11,2% | 67,7% | 51,8% |
| | *p-value* (teste-f) | 4,6E-10 | 8,8E-02 | 3,3E-06 | 1,9E-03 |
| EF | $R^2$ | 65,1% | 60,3% | 53,0% | 54,7% |
| | *p-value* (teste-f) | 1,5E-11 | 2,3E-06 | 1,4E-04 | 1,2E-03 |
| | Proportion to FP's $R^2$ | +10% | +438% | -22% | +5% |
| EFt | R2 | 66,1% | 60,3% | 53,0% | 54,7% |
| | *p-value* (teste-f) | 8,5E-12 | 2,3E-06 | 1,4E-04 | 1,2E-03 |
| | Proportion to FP's $R^2$ | +11% | +438% | -22% | +5% |

**Table X:** Justifications Of How The New Metrics Address The Critiques Presented In Section 2.2

| Criticism | Solution |
|---|---|
| Low representation | Each possible combination of the functional attributes considered in deriving the complexity in FPA is associated with a distinct value. |
| Functions with different complexities have the same size | Functionalities with different complexities, as determined by the number of functional attributes, assume a different size. |
| Abrupt transition between functional element ranges | By applying the formulas of calculation described in Table VII in Section 2.3.4, the variation in size is uniform for each variation of the number of functional attributes, according to its coefficients. |
| Limited sizing of high complexity functions | There is no limit on the size assigned to a function by applying the calculation formulas described in Table VII in Section 2.3.4. |
| Undue operation on ordinal scale | The metrics do not have a ordinal scale with finite values, but rather a quantitative scale with infinite discrete values, which provide greater reliability in operations with values. |
| Inability to measure changes in parts of the function | Enables the measurement of changes in part of a functionality considering in the calculation only the functional attributes impacted by the amendment. |

Given the observations listed above, we conclude that the metrics proposed, EF and EFt[21], have better correlation with effort in comparison to FPA for the analyzed data[22].

Table X contains the explanation of how the proposed metrics, EF and EFt, address the criticisms presented in Section 2.2.

## 2.5 ILLUSTRATION OF THE USE OF THE NEW METRICS IN IT GOVERNANCE

Kaplan and Norton [1992, p. 71] claim that what you measure is what you get. According to COBIT 5 [ISACA, 2012b, p. 13], governance aims to create value by obtaining the benefits through optimized risks and costs. In relation to IT governance, the metrics proposed in this paper not only help to assess the capacity of IT but also enable the optimization of its processes to achieve the results.

Metrics support the communication between the different actors of IT governance (see Fig. 3) by enabling the translation of objectives and results in numbers. The quality of a process can be increased by stipulating objectives and by measuring results through metrics [MOSER, 1996, p. 19]. So, the production capacity of the process of information systems development can be enhanced to achieve the strategic objectives with the appropriate use of metrics and estimates.

Software metrics contribute to the three IT governance activities proposed by ISO 38500, mentioned in Section 1.1: to assess, to direct and to monitor. These activities correspond, respectively, to the goals of software metrics mentioned in Section 2.1.1: to understand, to improve, and to control the targeted entity of a measurement.

Regarding the directions of IT area, Weill and Ross [2006, p. 188] state that the creation of metrics for the formalization of strategic choices is one of four management principles that summarize how IT governance helps companies achieve their strategic objectives. Metrics must capture the progress toward strategic goals and thus indicate if IT governance is working or not [WEIL; ROSS, 2006, p. 188].

Kaplan and Norton [1996, pp. 75-76] claim that strategies need to be translated into a set of goals and metrics in order to have everyone's commitment. They claim that the Balanced Scorecard (BSC) is a tool which provides knowled-

**Figure 3:**
Roles, activities and relationships of IT governance

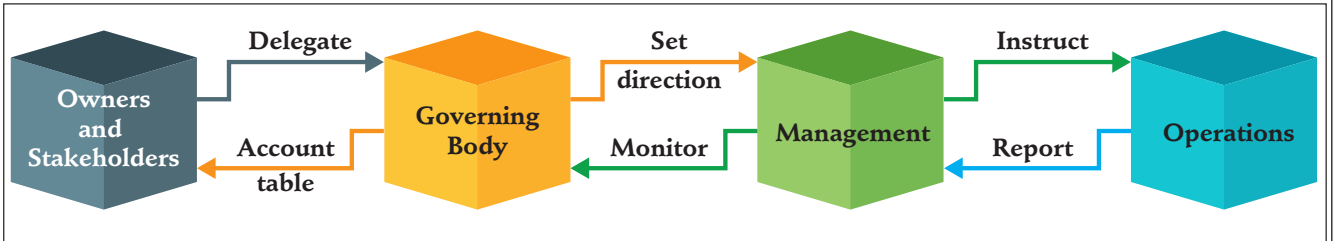

Adapted from ISACA [2012a, p. 24]

ge of long-term strategies at all levels of the organization and also promotes the alignment of department and individual goals with those strategies. According to ITGI [2007, p. 29], BSC, besides being a holistic view of business operations, also contributes to connect long-term strategic objectives with short-term actions.

To adapt the concepts of the BSC for the IT function, the perspectives of a BSC were re-established [VAN GREMBERGEN; VAN BRUGGEN, 1997, p. 3]. Table XI presents the perspectives of a BSC-IT and their base questions.

According to ITGI [2007, p. 30], BSC-IT effectively helps the governing body to achieve alignment between IT and the business. This is one of the best practices for measuring performance [ITGI, 2007, p. 46]. BSC-IT is a tool that organizes information for the governance committee, creates consensus among the stakeholders about the strategic objectives of IT, demonstrates the effectiveness and the value added by IT and communicates information about capacity, performance and risks [ITGI, 2007, p. 30].

Van Grembergen [2000, p.2] states that the relationship between IT and the business can be more explicitly expressed through a cascade of scorecards. Van Grembergen [2000, p.2] divides BSC-IT into two: BSC-IT-Development and BSC-IT-Operations. Rohm and Malinoski [2010], members of the Balanced Scorecard Institute, present a process with nine steps to build and implement strategies based on scorecard. Bostelman and Becker [1999] present a method to derive objectives and metrics from the combination of BSC and the Goal Question Metric (GQM) technique proposed by Basili and Weiss [1984]. This association between BSC and GQM is consistent to what ISACA [2010, p. 74] says: good strategies start with the right questions. The metric proposed in this paper can compose several indicators that can be used in BSC-IT-Development.

Regarding the activities of IT monitoring and assessment [ISO; IEC, 2008, p. 7], metrics enable the monitoring of the improvement rate of organizations toward a mature and improved process [RUBIN, 1993, p. 473]. Performance measurement, which is object of monitoring and assessment, is one of the five focus areas of IT governance, and it is classified as a driver to achieve the results [ITGI, 2007, p. 19].

To complement the illustration of the applicability of the new metric for IT governance, Table XII shows some indicators based on EF. The same indicator can be used on different perspectives of a BSC-IT-Development, depending on the targeted entity and the objective of the measurement, such as the following examples. The productivity of a resource (e.g., staff, technology) may be associated with the *Future Orientation* perspective, as it seeks to answer whether IT is prepared for future needs. The same indicator, if associated with an internal process, encoding, for example, reflects a vision of its production capacity, in the *Operational Excellence* perspective. In the *Customer Orientation* perspective, production can be divided by client, showing the proportion of IT production to each business area. The evaluation of the variation in IT production in contrast to the production of business would be an example of using the indicator in the *Contribution to the Business* perspective.

The choice of indicators aimed to encompass the five fundamental dimensions mentioned in Section

**Table XI:**
Perspectives Of A Bsc-It
Source: inspired in ITGI [2, p. 31]

| Perspective | Base question | BSC corporative perspective |
|---|---|---|
| Contribution to the business | How do business executives see the IT area? | Financial |
| Customer orientation | How do customers see the IT area? | Customer |
| Operational excellence | How effective and efficient are the IT processes? | Internal Processes |
| Future orientation | How IT is prepared for future needs? | Learning |

**Table XII:**
Description
Of Illustrative
Indicators

| Metric | Unit | Dimension | Description of the calculation for a system |
|---|---|---|---|
| Functional size | EF | Size | sum of the functional size of the functionalities that compose the system at the end of the period |
| Production in the period | EF | Effort | sum of the functional size of requests for inclusion, deletion, and change implemented in the period |
| Production on rework | EF | Rework | sum of the functional size of requests for deletion and change implemented in the period |
| Productivity | Functional Elements / Man–hour | Effort | sum of the functional size of requests implemented in the period / sum of the efforts of all persons allocated to the system activities in the period |
| Error density | Failures / Functional Element | Quality | number of failures resulting from the use of the system in a period / size of the system at the end of the period |
| Delivery speed | Functional Elements / Hour | Time | sum of the size of the features implemented in the period / elapsed time |
| Density of the expected benefit | $ / EF | Expected benefit | benefit expected by the system in the period / system size |

II.11: size, effort, time, quality, and rework. Another dimension was added: the expected benefit. According to Rubin [2003, p. 1], every investment in IT, from a simple training to the creation of a corporate system, should be aligned to a priority of the business whose success must be measured in terms of a specific value[23]. The dimension of each indicator is shown in the third column of Table XII.

Some measurements were normalized by being divided by the number of functional elements of the product or process, tactics used to allow comparison across projects and systems of different sizes. The ability to standardize comparisons, as in a BSC, is one of the key features of software metrics [HUFSCHMIDT, 2002, p. 493]. It is similar to normalize construction metrics based on square meter, a common practice [DEKKERS, 2002, p. 161].

As Dennis argues [2002, p. 302], one should not make decisions based on a single indicator, but from a vision formed by several complementary indicators. As IT has assumed greater prominence as a facilitator to the achievement of business strategy, the use of dashboards to monitor its performance, under appropriate criteria, has become popular among company managers [ISACA 2010, p. 74]. Abreu and Fernandes [2009, p. 167] propose some topics that may compose strategic and tactical control panels of IT.

Graph 3 illustrates the behavior of the indicators shown in Table XII with annual verification for hypothetical systems. The vertical solid line indicates how the indicator to the system was in the previous period, allowing a view of the proportion of the increasing or decreasing of the values over the period. In the productivity column (column 4), a short line at its base indicates, for example, a pattern value obtained by benchmark. The vertical dashed line metric associated with the production in the period (2) indicates the target set in the period for each system: system A reached it, system D exceeded it, and systems B and C failed.

In one illustrative and superficial analysis of the indicators for system C, one can associate the cause of not achieving the production goal during that period (2) with the decrease of the delivery speed (6) and with the increase of the production on rework (3), resulted, most likely, from the growth in the error density (5). The reduction on the delivery speed (6) which can be associated with decreased productivity (4) led to a low growth of the functional size of the system (1) during that period. These negative results led to a decrease in the density of the expected benefit (7).

Graph 3 represents an option of visualization of the governance indicators shown in Table XII: a multi-metrics chart of multi-instances of a targeted entity or a targeted attribute. The vertical column width is variable depending on the values of the indicators (horizontal axis) associated with the different instances of entities or attributes of interest (vertical axis). The same vertical space is allocated for each entity instance. The width of the colored area, which is traced from the left to the right, indicates graphically the value of the indicator for the instance.

In the hands of the governance committee, correct indicators can help senior management, directly or through any governance structure, to identify how IT

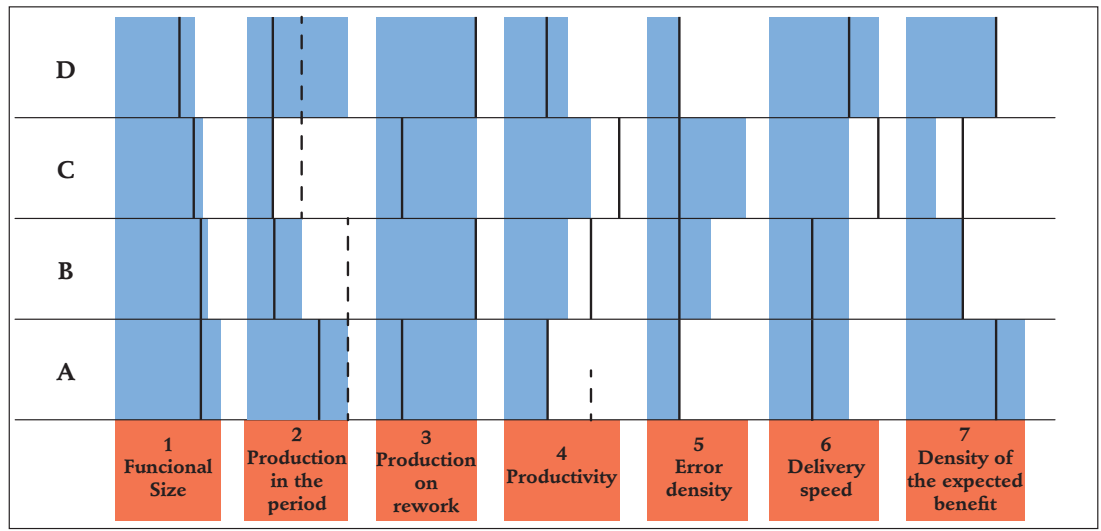

**Graph 3:**
Annual indicators of systems A, B, C and D

Columns labeled: 1 Funcional Size; 2 Production in the period; 3 Production on rework; 4 Productivity; 5 Error density; 6 Delivery speed; 7 Density of the expected benefit management is behaving and to identify problems and the appropriate course of action when necessary.

## 3. FINAL CONSIDERATIONS

The five specific objectives proposed for this work in Section 1.4 were achieved, albeit with limitations and with possibilities for improvement that are translated into proposals for future work.

The main result was the proposition of a new metric EF and its submetric EFt. The new metrics, free of some deficiencies of the FPA, metrics taken as a basis for their derivation, reached a higher correlation with effort than the FPA metric, in the context of the analyzed data.

The paper also illustrated the connection between metrics and IT governance activities, either in assessment and monitoring, through use in dashboards, or in giving direction, through use in BSC-IT.

There are possibilities for future work in relation to each of the specific objectives.

Regarding the conceptualization and the categorization of software metrics, a comprehensive literature research is necessary to the construction of a wider and updated categorization of software metrics.

Regarding the presentation of the criticisms to FPA, only the criticisms addressed by the new proposed metrics were presented. Research in the theme, as a bibliographic research to catalog the criticisms, would serve to encourage other propositions of software metrics.

Regarding the process of creating the new metric, it could be improved or it could be applied to other metrics of any area of knowledge based on ordinal values derived from tables of complexity as FPA (e.g., metric proposed by KARNER [1993]: Use Case Points). Future works may also propose and evaluate changes in the rules and in the scope of the EF. The creation process could be improved for example, by treating differently the unlimited ranges section (2.3.2). Weights could be attributed to sizes of the limited ranges, for example, according to the proportion of the functions that integrate such ranges in a sample with functionalities from several systems.

Regarding the evaluation of the new metric, the limitation in using data from only one organization could be overcome in new works. Practical applications of the metric could also be illustrated, for example, in contracts with an incremental delivery process. New works could compare the results of EF with the EFt submetric as well as compare both with other software metrics. Different statistical models could be used to evaluate its correlation with effort even in specific contexts (e.g., development, maintenance, development platforms). We expect to achieve a higher correlation of the new metric with effort in agile methods regarding to the FPA, considering its capacity of partial functionality sizing. (6th criticism in Section 2.2.)

Regarding the connection with IT governance, a work about the use of metrics in all IT governance activities is promising. The proposed graph[25] for visualization of multiple indicators of multiple instances through columns with varying widths along their length could also be standardized and improved in future work.

A suggestion for future work is noteworthy: the definition of an indicator that shows the level of maturity of a company regarding to the use of metrics in IT governance. Among other aspects, it could consider in the composition of the indicator, the following are noteworthy: the breadth of the entities evaluated (e.g., systems, projects, processes, teams), the dimensions treated (e.g., size, rework, quality, benefits) and the effective use of the indicators (e.g., monitoring, direction).

Finally, we expect that the new metric EF and its submetric EFt help increase the contribution of IT to the business in an objective, reliable, and visible way.

## NOTES

1. A version of this paper, in English, not including all the content here, was presented at the XXVII SBES (Brazilian Symposium on Software Engineering) promoted by SBC (Brazilian Computer Society) and was published in the IEEE Xplore: M. V. B. D. Castro and C. A. M. Hernandes, "A Metric of Software Size as a Tool for IT Governance", *Software Engineering (SBES), 2013 27th Brazilian Symposium on*, Brasilia, 2013, pp. 99-108. . doi: 10.1109/SBES.2013.13..

2. In its version, 4.3.1 (IFPUG, 2010), appendix C, there is a possibility of adjusting the functional size of a factor that reflects an assessment of the system in relation to 41 general non-functional characteristics. According to Fenton and Pfleeger (1998, p. 262), the determination is subjective and according to Kemerer (1987, p. 9), the adjustment does not increase the correlation of metric with effort. This part was separated from the standard rule of function points because FPA is an ISO standard of functional metric only without application of the adjustment.

3. There are several rulings on the subject: 1.782/2007, 1.910/2007, 2.024/2007, 1.125/2009, 1.784/2009, 2.348/2009, 1.274/2010, 1.647/2010, all of the Plenary of the TCU.

4. Also known as CIO – Chief Information Officer.

5. Kitchenham et al (1995) present a framework for the software metric in which they list the concepts associated with the formal model on which the metric is based (e.g. type of scale used).

6. The overview presented results from the experience with the FPA of Marcus, one of the authors. In 1993 he coordinated a program for implementation of the use of FPA in the development area of the Superior Labor Court (TST). At the TCU, he also works with metric.

7. Functional requirements are only one dimension of several impacting the effort. All of them have to be taken into account in estimates. Estimates and non-functional requirements evaluations are not the goal of this paper.

8. Functionalities with a very low level of complexity are also not dimensioned appropriately for FPA because they take on the minimum value when they should take on an even smaller value.

9. Because they are concepts that are widely known by the measurers, it is expected that the new metric will be accepted among the professionals of the field.

10. In this paper, these attributes correspond to the concept of functional elements, name of the proposed metric.

11. The alternative of attributing a third range to the sum of elements of the two first ranges was also assessed. However, this approach was less efficiant in the correlation with effort, for all data evaluated.

12. The choice was to distinguish the EFt metric for application in cases where the effort to treat data structures (Efd) is not the object of assessment or contract. Although it has not been assessed, the EFd submetric has its role in translating the structural complexity of data of an application

13. Kemerer (1987, p. 428) justified the use of linear regression as a means to evaluate the correlation of FPA metric with effort.

14. The agency that provides the data informed that each system was implemented in one language only: Java, DotNet or Natural.

15. The order of the systems follows the criterium of the quantity of OS.

16. A logistic non-linear regression was also carried out, with a constant, using the Gretl software, a free open code tool (http://www.simula.no/BESTweb), created as a result of the Jörgensen and Shepperd research (2007). However, the R2 factor proved that this alternative was worse than the linear regression for all metrics and, therefore, the nonlinear model correlation was discarded. R2 nonlinear regression reached the following values: (system R2_APF, R2_PM, R2_EF, R2_EFt) - (M, 0.316, 0.470, 0.434, 0.426); (B 0.013, 0.313, 0.442, 0.443); (C, 0.327, 0.262, 0.16, 0.152) and (D, 0.02, 0.127, 0.087, 0.087).

17. That is, the line goes through the origin of the axes.

18. This restriction is justified, for example, by the information given by the agency providing the data that the development language is only one per system and that the technical team is, as a general rule, also the same per system. The language and the team are factors that influence effort. The factors that influence effort and the degree of this correlation were discussed in several articles. For more details on the topic, we suggest accessing the articles in the base BestWeb (http://www.simula.no/BESTweb), created as a result of the Jörgensen and Shepperd research (2007).

19  In order to consider a correlation as statistically significant at a X% level of reliability, the p-value must be smaller than a 1 – X (ORLOV, 2996, p. 11). For a 95% level, the p-value needs to be smaller than 0,05.

20  The criterion used to consider the correlation C1 superior to correlation C2: C1 being significant and C2 not significant or, if both are significant, C1 having a larger R2 than C2.

21  We notice a greater correlation of the EFt metric in relation to EF in the H system, the only system that enabled a difference in the result of the two metrics by presenting commands related to the alteration of logic files in its OS. Thus, we notice a trend that is favorable for submetric EFt in relation to EF, reinforcing the hypothesis that submetric EFd which makes up the EF metric does not impact the coding and testing effort, tasks dealt with in the OS that were assessed.

22  A comparison between the correlation of the EF metric and the correlation of the PM metric (NESMA) was not the objective of the work. However, since the data also provided measures in PM, a metric used in contract of the data provider, an assessment of the PM metric was also carried out.

|    | System | H | B | C | D |
|---|---|---|---|---|---|
| PM | $R^2$ | 67,0% | 48,1% | 63,7% | 60,4% |
|    | p-value (teste-f) | 4,7E-12 | 6,7E-05 | 1,1E-05 | 4,8E-04 |
|    | comparison $R^2$ PF | +13% | +329% | -6% | +17% |

**Comments**

1. all correlations were considered significant at a confidence level of 95%;

2. as the new metric, PM achieved average correlations (0.8> R-squared> 0.5) for the four systems and a superior result than the FPA in the same three systems (H, B and D);

3. in the H system there was a equivalence of PM correlation with EFT correlation, with 0.9%, as the difference between the two correlations;

4. PM correlations were higher in 2 systems in relation to the new metric (C and D) and lower in a system (B).

5. in the context of the data assessed, the numbers show a slight superiority of PM in relation to the new metrics. However, we can foresee a potential of better results for EF, since PM has the first five flaws mentioned in section 2.2.

6. PM has another conceptual flaw: in contrast with the FPA and EF, it dimensions in a different way software that is under development and software in maintenance, with different adjustment factors. The maintenance cost can actually be different with relation to development. Since it can also be different due to other factors (e.g. language used, methodology applied). Factors that do not alter the size of software, but the cost. Bringing concern with cost to the size of the software, as PM does, does not seem to be the best option because these are different concepts.

7. another disadvantage of PM in relation to the proposed metric is the counting cost. PM requires the length of the functionality before the maintenance, and no extra effort is required for the new metrics.

23  The scope of the work does not include investigating the concepts and processes associated to determining the value of a function or of a system or IT area. It is a complex topic that is still incipient.

24  The maximum value for each indicator in the period was associated to the maximum width defined for the column. The widths of the colored areas of the other systems were derived by using a simple rule of three.

25  At http://learnr.wordpress.com/ (access on Nov 4. 2012) there is a graph, which is functionally similar to what is proposed: *heatmap plotting*, however it is different as to format and possibilities of evolution. Since a similar graph was not found, the assumption is that it is a new format to visualize the behavior of multiple indicators in multiple instances in columns with variable widths in their extension (MIMICoVaWE - Multiple Indicators about Multiple Instances through Columns with Varying Widths along their Extension). Two examples among the several possible evolutions for a graph: an exchange in the position between the metrics and the instances, with these passing through the horizontal axis, a variation in the color tone of the cell that relates a metric to an instance according to some criterion (e.g. a regarding achievement of a specified goal).

## REFERÊNCIAS